\documentclass[aps, prd, tightenlines, notitlepage, superscriptaddress, nofootinbib, preprintnumbers, floatfix,showkeys,11pt,altaffilletter]{revtex4-2}

\usepackage{upgreek}

\usepackage[official]{eurosym}
\usepackage[normalem]{ulem}
\usepackage{amstext}
\usepackage{graphicx}
\graphicspath{{figures/}}
\usepackage{url}
\usepackage{color}
\usepackage{ulem}
\usepackage[version=4]{mhchem}
\usepackage[utf8]{inputenc}
\usepackage{fontawesome}
\usepackage{yfonts}
 
\pdfoutput=1
\usepackage{textcomp}
\usepackage{comment}
\usepackage{yfonts}
\usepackage{epsfig,amsfonts,mathrsfs,graphicx,color,slashed,multirow}
\usepackage{latexsym,graphicx,slashed,color,enumerate,url,cancel,gensymb}
\usepackage{textcomp}

\usepackage[x11names]{xcolor}
\usepackage[colorlinks,pdfstartview=FitV,breaklinks=true]{hyperref}
\usepackage{booktabs}
\usepackage{adjustbox}
\usepackage{textgreek} 
\usepackage{caption, subcaption} 
\usepackage{float}

\usepackage{lmodern}
\usepackage{ae,aecompl}
\usepackage{appendix}


\definecolor{persiangreen}{rgb}{0.0, 0.65, 0.58}
\definecolor{mediumpersianblue}{rgb}{0.0, 0.4, 0.65}

\makeatletter
    \newcommand{\colorboxed}[3][white]{\fcolorbox{#2}{#1}{\m@th$\displaystyle#3$}}
\makeatother

\AtBeginDocument{\hypersetup{citecolor=mediumpersianblue,linkcolor=mediumpersianblue,urlcolor=mediumpersianblue}}
\usepackage{appendix}

\begin{document}

\title{{\LARGE Observing neutrinos from failed Supernovae at LNGS}}
\author{Giulia Pagliaroli}
\email{giulia.pagliaroli@lngs.infn.it}
\affiliation{Istituto Nazionale di Fisica Nucleare (INFN), Laboratori Nazionali del Gran Sasso, 67100 Assergi, L’Aquila (AQ), Italy
}
\author{Christoph A. Ternes}
\email{christoph.ternes@lngs.infn.it}
\affiliation{Istituto Nazionale di Fisica Nucleare (INFN), Laboratori Nazionali del Gran Sasso, 67100 Assergi, L’Aquila (AQ), Italy
}

\begin{abstract}
We discuss the possibility to observe neutrinos emitted from a failed core collapse Supernova in the various experiments at Laboratori Nazionali del Gran Sasso. We show that the veto regions of dark matter and neutrinoless double beta decay experiments can be used as a network of small detectors to measure Supernova neutrinos. In addition we show that this network can measure very precisely the moment of black hole formation, which can be then used in the nearby VIRGO detector and future Einstein Telescope to look for the gravitational wave counterpart to the neutrino signal.
\end{abstract}
\maketitle

\section{Introduction}
\label{sec:intro}

When stars with masses $M>8M_\odot$ burn all their fuel the radiation pressure in the iron core region becomes not sufficient to balance the gravitational force. The iron core collapses until the central density overcomes the nuclear density. At this point the outer part of the core bounces off the inner compact region. The bounce induces a shock wave that, after a stalled phase, can be revived by neutrinos providing the final Supernova explosion \cite{Burrows:2020qrp}.
The Core Collapse Supernova (CCSN) releases an energy of several $10^{53}$ ergs and neutrinos carry away most of this energy ($99\%$) over a time interval of about 10~s \cite{Janka2017}. So far, only the Supernova SN1987A has been observed in neutrinos, due to the low CCSN rate within the Milky Way (viz. $R_{\text{MW}} = 1.63\pm0.46$ per century~\cite{Rozwadowska:2020nab}). However, the small data-set of 24 events observed by the Kamiokande-II~\cite{Kamiokande-II:1987idp,Hirata:1988ad}, Baksan~\cite{Alekseev:1987ej} and IMB detectors~\cite{Bionta:1987qt} testifies our capability to detect this emission and to use neutrino data to constrain CCSN emission models~\cite{Pagliaroli:2008ur}. 
Current and future detectors would see thousands of events in neutrinos from a Galactic CCSN~\cite{SNEWS:2020tbu}
providing a fast alert to optical telescopes to catch the upcoming electromagnetic emission and shock breakout~\cite{Nakar:2010tt}.

In the case of more massive stars, with $M\gtrsim 20M_\odot$~\cite{Bahcall:1982fx}, it is possible that the explosion mechanism fails and the accretion of matter on the central proto-neutron star generates a black hole (BH) very shortly after the core bounce. In this scenario, called failed SN,
neutrino and gravitational wave (GW) emissions stop abruptly at the same time \cite{OConnor:2010moj,Baumgarte:1996xz}.
This astrophysical event is expected to emit only a very faint electromagnetic signal from the disappearing progenitor star \cite{Adams:2016ffj}, so that the observation of neutrinos and gravitational waves can be the only way to hunt for the BH formation. 

Neutrinos from Supernovae can be detected in different types of detectors with different detection channels and neutrino flavor sensitivities~\cite{Scholberg:2012id}. They can be looked for in Cherenkov detectors~\cite{AMANDA:2001htp,IceCube:2011cwc,KM3NeT:2021moe,Super-Kamiokande:2022dsn,IceCube:2023ogt}, liquid scintillator detectors~\cite{MACRO:2004fzv,SNO:2010noh,LVD:2014uzr,Novoseltsev:2019gdt,Rumleskie:2020iip,NOvA:2021zhv,KamLAND:2022sqb}, in lead-based detectors~\cite{GalloRosso:2020qqa}, in noble gas detectors~\cite{DUNE:2020zfm} or also in Dark Matter detectors~\cite{Lang:2016zhv}. 

The Gran Sasso National Laboratory or Laboratori Nazionali del Gran Sasso (LNGS) hosts at present the LVD experiment \cite{LVD:2014uzr}, aimed at CCSN neutrino detection, and many other experiments dedicated to different physics topics, nowadays with an important focus on neutrinoless double beta decay and direct detection of dark matter. These experiments use large water tanks as veto regions against different types of backgrounds, such as cosmic muons or natural radioactivity in the surrounding cave material. These veto regions can be used as a network of Water Cherenkov detectors to look for neutrinos from CCSNe. 

The aim of this paper is twofold: On one hand, we suggest to exploit the veto regions of detectors, generally dedicated to other physics topics, as a free-of-charge network of small detectors for neutrinos from CCSNe; on the other hand, we discuss the possibility of neutrino detectors to identify the time of black hole formation in very massive stars providing an important timing to look for the gravitational counterpart of such events. 
We show that, in case of a failed Supernova the LNGS infrastructure can provide precise information on the time when the BH forms. This timing could be crucial for the search for the associated gravitational waves. In the absence of the electromagnetic counterpart, LNGS is the only facility that can provide this timing in real time without the need to know the sky position of the SN, thanks to the negligible time of fly to the VIRGO gravitational wave interferometer \cite{VIRGO:2014yos} and to the future Einstein Telescope \cite{Punturo:2010zz}, located in Pisa and in Sardinia \cite{Naticchioni:2024npy}, respectively. 

Our paper is structured as follows: We discuss in Section~\ref{sec:emission} the neutrino emission models under consideration in this work. Section~\ref{sec:det} contains information on the current and future LNGS-detectors and the calculation of the event rates used in our analysis. We discuss our results in Section~\ref{sec:results} and we draw our conclusions in Section~\ref{sec:conclusions}.

\section{Neutrino emission from failed Supernovae}
\label{sec:emission}

\begin{figure}
\centering
\includegraphics[width=0.49\textwidth]{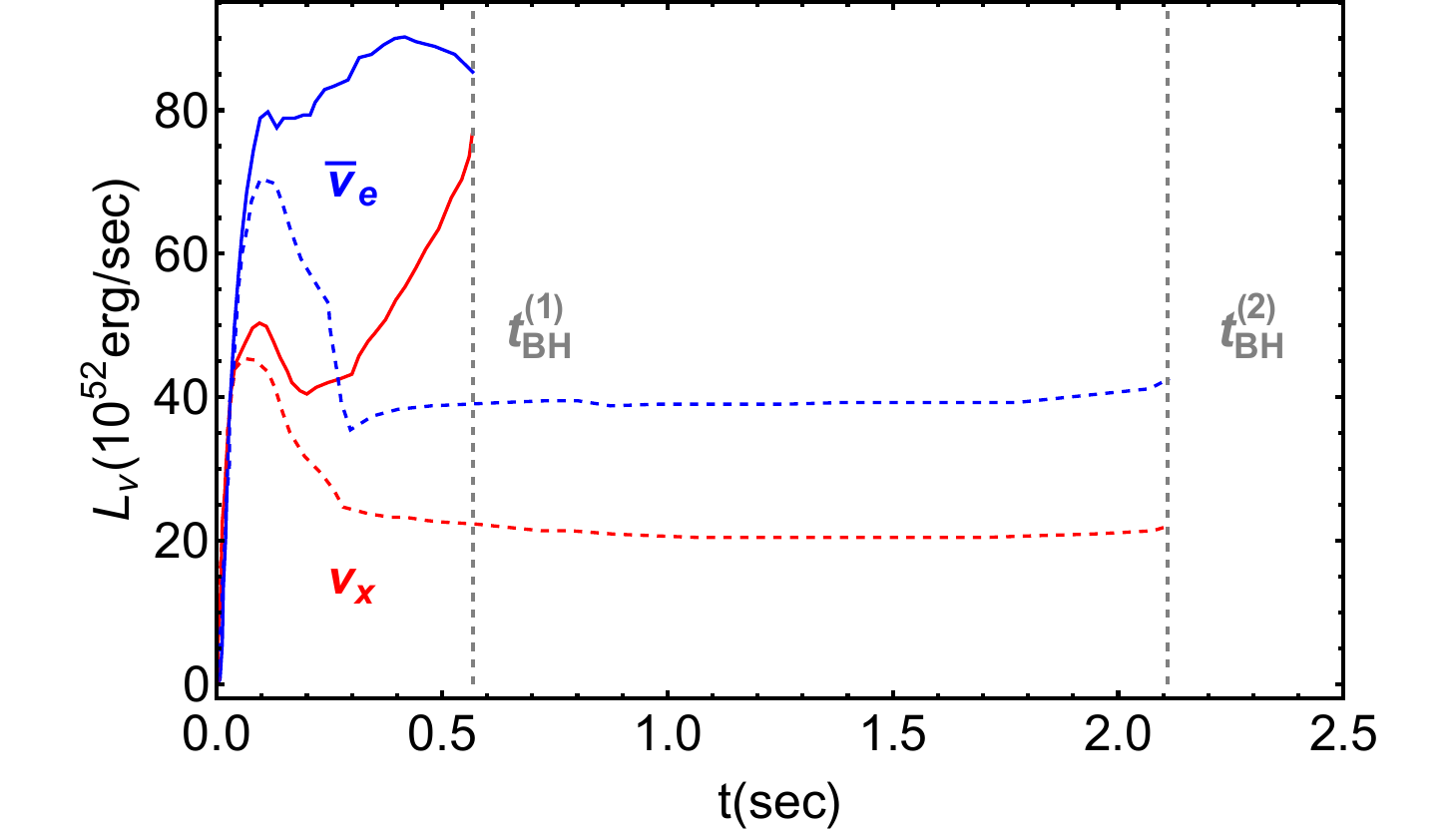}
\includegraphics[width=0.49\textwidth]{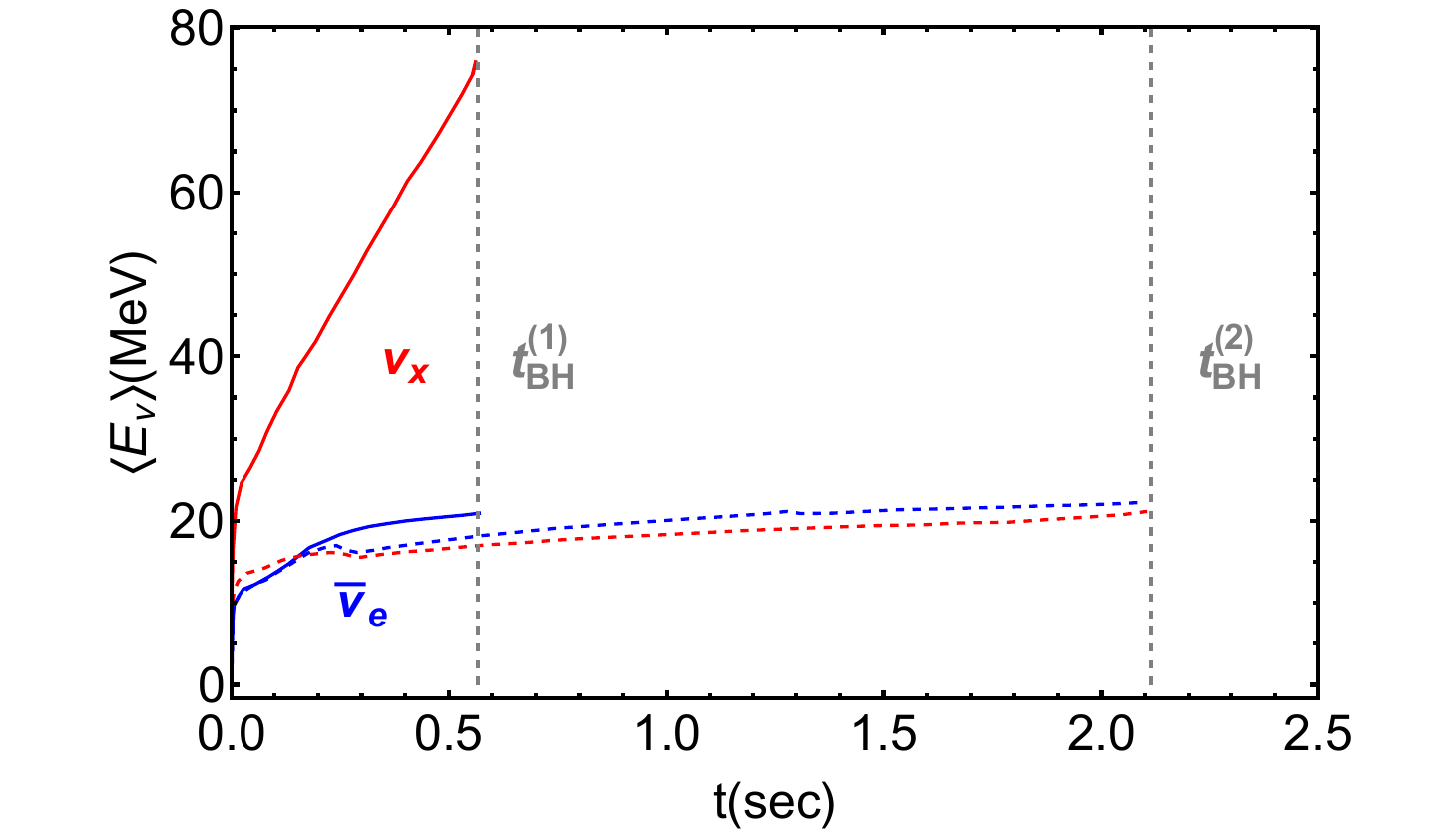}
\caption{Neutrino luminosity $L_{\nu_\alpha}$ and average energy $\langle E_{\nu_\alpha}\rangle$ as a function of time after the core bounce for Model 1 (solid lines) and Model 2 (dashed lines) discussed in the text. With a dashed gray line we highlight the time of the black hole formation in both cases.}
\label{fig:luminosity}
\end{figure}

In order to investigate the capability of the LNGS detectors to provide a good timing measurement of the BH formation in failed SNe we consider the two different models reported in Fig.~15 of Ref.~\cite{Mirizzi:2015eza}. In both cases, the progenitor, a star of 40 $M_\odot$, fails to explode and forms a central BH. 
The two models, both using the ls221 equation of state \cite{Lattimer:1991nc}, differ with respect to the duration of their evolution until BH formation. In Model 1 \cite{Woosley:1995ip}, the BH formation occurs $t_{\text{BH}}^{(1)}=0.568$~s after the bounce, whereas for Model 2 \cite{Woosley:2002zz} the gravitational instability occurs at a later time of $t_{\text{BH}}^{(2)}=2.113$ s after the bounce thanks to a lower mass accretion rate of the collapsing stellar core.
In Fig.~\ref{fig:luminosity}, we show the neutrino luminosity $L_{\nu_\alpha}$ and average neutrino energy $\langle E_{\nu_\alpha}\rangle$ as a function of time for both models. The origin represents the time of the core bounce inside the CCSN. In both cases the neutrino emission terminates abruptly at the time of BH formation $t_{\text{BH}}$. Following Ref.~\cite{Keil:2002in}, the differential neutrino flux per flavor is defined as
\begin{equation}
\frac{d\Phi_\alpha^0}{dE_{\nu_\alpha}dt}=\frac{1}{4\pi d^2} \frac{L_{\nu_\alpha}(t)}{\langle E_{\nu_\alpha}(t)\rangle}\phi_\alpha(E_{\nu_\alpha},t),
\end{equation}
where $d$ is the SN distance and the normalized energy spectrum is parameterized as:
\begin{equation}
\phi_\alpha(E_{\nu_\alpha},t)=\langle E_{\nu_\alpha}(t)\rangle \frac{1+\xi_\alpha}{\Gamma(1+\xi_\alpha)}\left(\frac{E_{\nu_\alpha}}{\langle E_{\nu_\alpha}(t)\rangle}\right)^{\xi_\alpha}\exp\left(-\frac{(1+\xi_\alpha)E_{\nu_\alpha}}{{\langle E_{\nu_\alpha}(t)\rangle}}\right),
\end{equation}
where the "pinching parameter" is set to $\xi_\alpha=3$.
Due to neutrino oscillations, the flux of electronic antineutrinos arriving at the detector is a combination of initial $\bar{\nu}_e$ and $\bar{\nu}_x$ at the emission, see e.g. Ref.~\cite{Mirizzi:2015eza} for a deep discussion.
Assuming normal neutrino mass ordering (currently moderately preferred at 2.7$\sigma$ over the inverted one~\cite{deSalas:2020pgw,Gariazzo:2022ahe}), neglecting collective effects and considering a static supernova matter profile, the electronic antineutrino flux which arrives at the detector is given by 

\begin{equation}
\Phi_{\bar{\nu}_e}=\cos^2\theta_{12}\Phi_{\bar{\nu}_e}^0+\sin^2\theta_{12}\Phi_{\bar{\nu}_x}^0.   
\end{equation}
Here we will concentrate on the electronic antineutrinos being the inverse beta decay the most relevant detection channel for the experiments under consideration. Note, however, that with other channels also different neutrino flavors might be observable.

\section{The LNGS detectors}
\label{sec:det}

At present the LNGS facility hosts the Large Volume Detector (LVD) \cite{LVD:2014uzr} that is a 1~kton liquid scintillator detector dedicated to search SN neutrinos and several detectors which are already taking data or still under construction and are dedicated to direct detection of dark matter or $0\nu\beta\beta$ decay search. Remarkably, several of these detectors have a veto region sensitive to SN neutrinos. In particular, the veto regions of Cryogenic Observatory for SIgnals seen in Next-generation Underground Searches (COSINUS) \cite{COSINUS:2021bdj}, the Large Enriched Germanium Experiment for Neutrinoless double beta Decay (LEGEND-200) \cite{Burlac:2021sbu} and XENONnT \cite{XENON:2023cxc} are water tanks with different shapes containing about $270$~tons, $590$~tons and $700$~tons of water, respectively.  
These are the experiments which are currently present at LNGS. We will also consider the future experiments LEGEND-1000~\cite{LEGEND:2021bnm} and DARWIN~\cite{DARWIN:2016hyl} (which might be hosted at LNGS) containing $980$~tons and $2140$~tons of water, respectively. These are the successor experiments to LEGEND-200 and XENONnT.
\begin{figure}[t]
\centering
\includegraphics[width=0.49\textwidth]{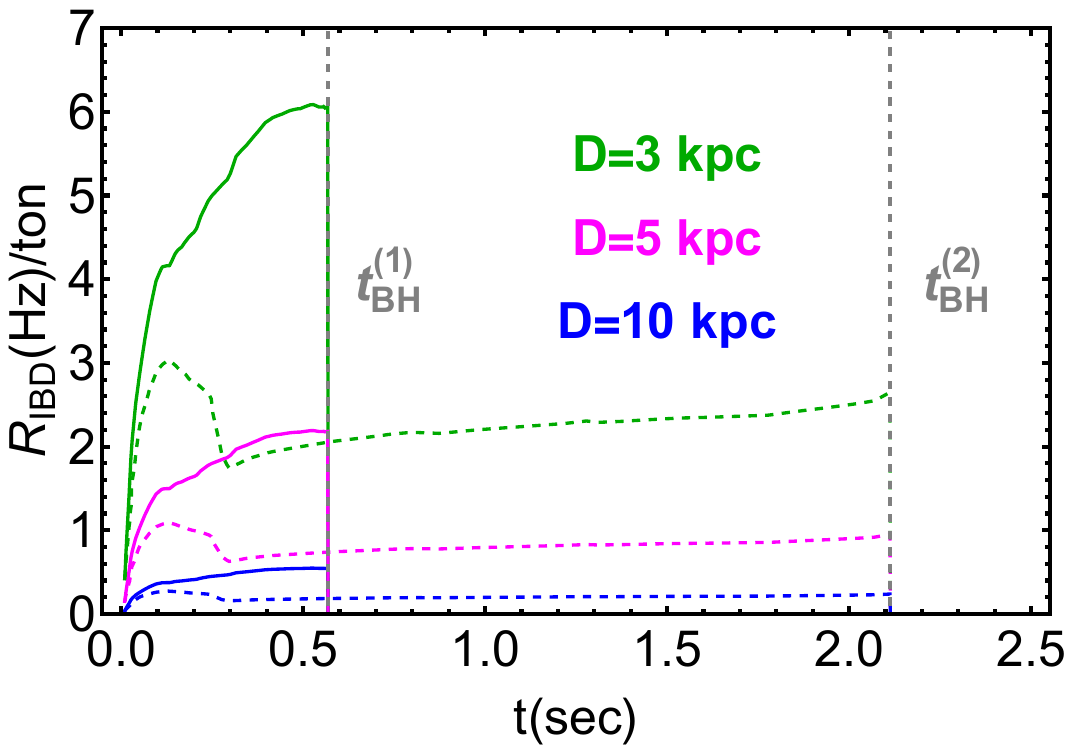}
\caption{Expected rate in Hz of IBD events in $1$ ton of water for different SN distances in kpc for emission Model 1 (solid lines) and Model 2 (dashed lines).}
\label{fig:rateIBD}
\end{figure}
The main interaction process for SN neutrinos in water and in liquid scintillators is the inverse beta decay (IBD) interaction of electronic antineutrinos with free protons.
The IBD event rate is given by 
\begin{equation}
R_{\text{\tiny IBD}}(t)=N_p\int_{E_{\text{\tiny thr}}}dE_{\nu}\sigma_{\text{\tiny IBD}}(E_\nu)\Phi_{\bar{\nu}_e}(E_\nu,t)\,,
\end{equation}
where $N_p$ is the number of free protons inside the detector, $\sigma_{\text{\tiny IBD}}$ is the IBD cross section taken from Ref.~\cite{Strumia:2003zx} and $\Phi_{\bar{\nu}_e}(E_\nu,t)$ has been discussed in Sec.~\ref{sec:emission}. 
As shown in Fig.~\ref{fig:rateIBD}, this rate can be very large (several kHz for the LNGS detectors under consideration) for a nearby failed SN, especially in the case of fast BH formation. This information should be considered in future experiments or the ones which are still under construction, in order to define the best data-acquisition rate enabling the goal proposed in this work.
The total number of IBD events expected as a function of the SN distance is reported in Fig.~\ref{fig:CCSNpresent} for the different detectors, each represented with a different color, and for the combined LNGS sensitivity in black. In the left panel we show the sensitivity of the currently present experiments, whereas in the right panel we show the one for future experiments. In each panel the solid (dashed) lines correspond to emission Model 1 (Model 2), see Sec.~\ref{sec:emission}. The lines labeled as "LNGS" show the expectations for the combined configuration where all the detectors are considered as different modules of a single big detector.  
We also show over-imposed the probability density distribution (PDF) of core-collapse SNe as a function of distance. This blue shadowed curve is obtained by considering the normalized Milky Way distribution of supernova remnants~\cite{Lorimer:2006qs} and by adding two Gaussian contributions for the Large and Small Magellanic Clouds centered at $d=49.5$~kpc and $62.8$~kpc, respectively. The width of each Gaussian accounts for the radial and angular size of both Galaxies\footnote{Taken from the NASA/IPAC Extragalactic Database (NED).} while the normalization is $11\%$ of the Milky Way contribution~\cite{Maoz:2010pz}. 
As can be easily observed the PDF shows a broad peak around $d=10$~kpc. For this reason we select this SN distance to discuss subsequent results. 
The number of expected IBD events for each detector and for a SN at $d=10$~kpc is reported in the first column of Tab.~\ref{tab:results} where we assumed an energy threshold of $4$~MeV for LVD~\cite{Vigorito:2021bbi} and $7$~MeV for the veto regions.
From Fig.~\ref{fig:CCSNpresent} it is evident that both the present configuration of LNGS detectors and the future one will be sensitive to measure SN neutrinos coming even from the Small Magellanic Cloud. 
As a final remark, it should be noted that the intrinsic modular configuration of all veto regions (and LVD) assures a very high duty cycle. Even if a single experiment was in maintenance during a SN, the other experiments could observe neutrinos individually.

\begin{figure}[t!]
\centering
\includegraphics[width=0.49\textwidth]{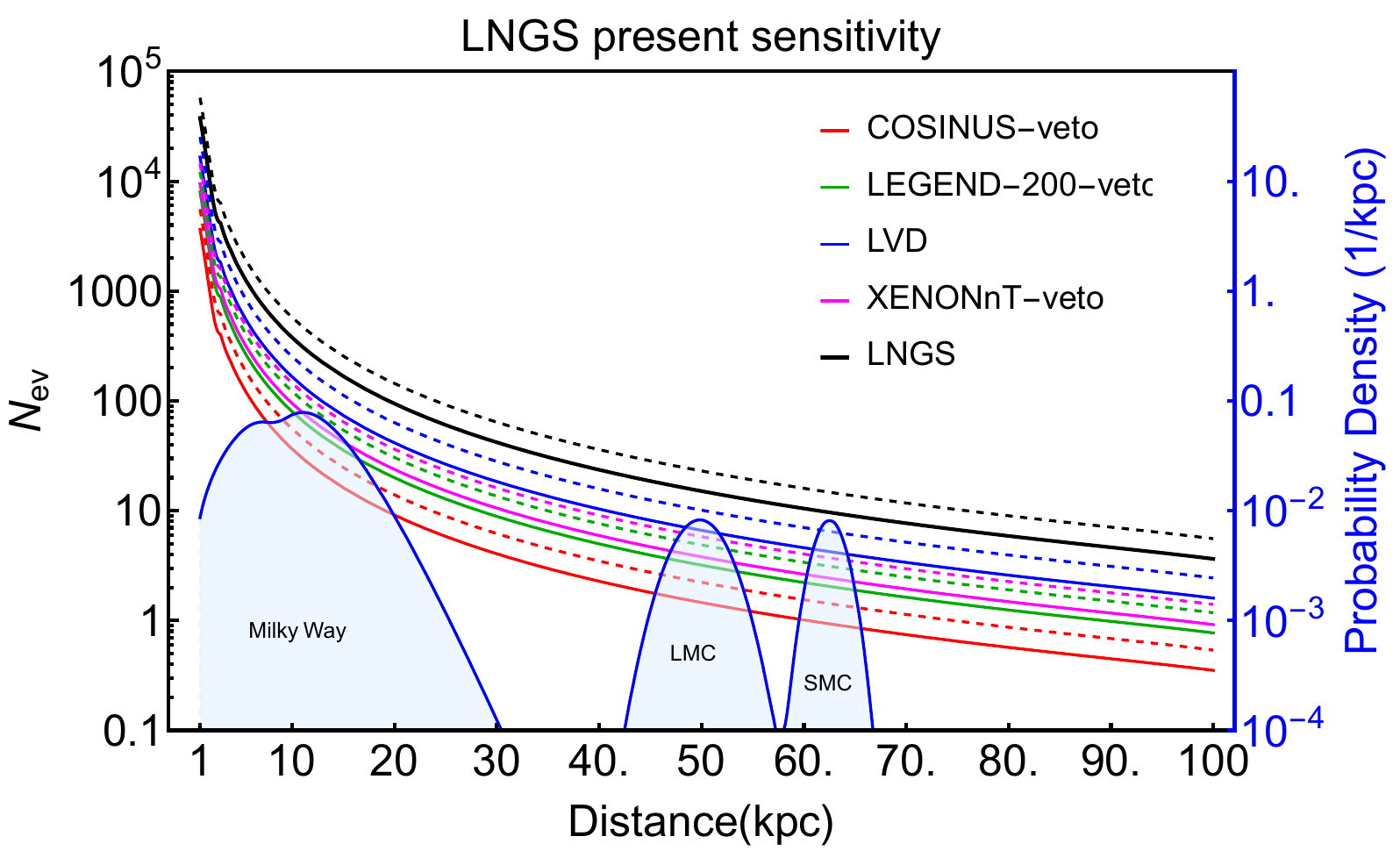}
\includegraphics[width=0.49\textwidth]{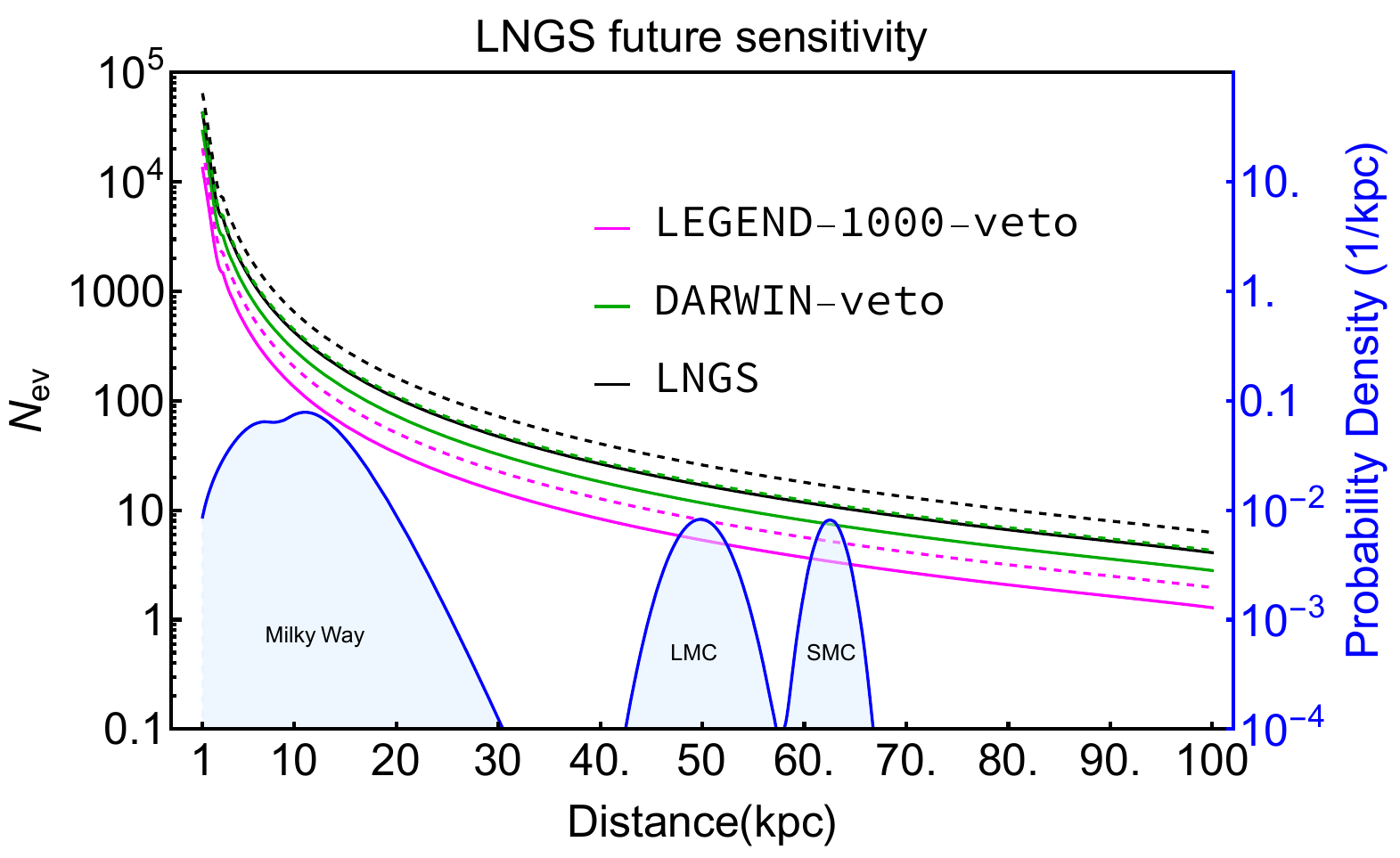}
\caption{Number of IBD events expected in the veto regions of current (left) and future (right) LNGS detectors as a function of the Supernova distance in kpc for Model 1 (solid lines) and Model 2 (dashed lines). The number of IBD events expected in LVD is also reported with a blue line in the left panel for comparison. The black lines provide the global LNGS sensitivity in this detection channel. We also show the probability density distribution (blue shadowed area) for a supernova explosion in the Milky Way and in the Magellanic Clouds, which has to be read with respect to the right (blue) axis.}
\label{fig:CCSNpresent}
\end{figure}

\section{Results}
\label{sec:results}

\begin{table}
\centering
    \begin{tabular}{|l|c|c|c|c|}
    \hline
       Detector & 
       $N_{\text{\tiny IDB}}$ & $t^{1}\pm\delta t^{1}$ [s] & $t^{\text{\tiny last}} \pm\delta t^{\text{\tiny last}} $ [s] & $1/\xi$ [s]\\
       \hline
         LVD & $303$ (527) & $0.018\pm 0.007$ ($0.019\pm 0.009$) & $0.566\pm 0.002$ ($2.109 \pm 0.003$) & 0.002 (0.004)\\
         COSINUS-veto& 66 (116) & $0.04\pm0.02$ ($0.04\pm0.02$) & $0.563\pm0.005$ ($2.10\pm 0.01$) & 0.008 (0.018)\\
         Legend200-veto& 145 (253) & $0.02\pm0.01$ ($0.03\pm0.01$) & $0.565\pm 0.003$ ($2.107\pm 0.007$) & 0.004 (0.008)\\
         XENONnT-veto& 173 (301) & $0.02\pm 0.01$ ($0.02\pm 0.01$)  & $0.565\pm0.003$ ($2.107\pm 0.005$) & 0.003 (0.007)\\
         \hline
         Legend1000-veto& 242 (530) &$0.019\pm 0.008$ ($0.021\pm 0.009$)& $0.566\pm0.002$ ($2.108\pm 0.005$) & 0.002 (0.005)\\
         DARWIN-veto& 530 (922) & $0.014\pm0.006$ ($0.013\pm0.006$) & $0.5672\pm0.0009$ ($2.111\pm0.003$) & 0.001 (0.002)\\
         \hline
    \end{tabular}
    \caption{The table shows part of the results obtained in our analysis. The numbers outside (inside) the parenthesis correspond to emission Model 1 (Model 2). In the first column we show the number of expected IBD events in the each detector. The quantity $t^1$ is the relative time between the time of the bounce and the time of the first neutrino detection averaged over 100 MC simulations. The statistical fluctuation of this quantity is labeled $\delta t^1$. The quantity $t^{\text{\tiny last}}$ is the average relative time between the last event and the time of the bounce and $\delta t^{\text{\tiny last}}$ is its statistical fluctuation. Finally the last column shows the inverse of the observed event rate, which is equivalent to the uncertainty of the time of BH formation.}
    \label{tab:results}
\end{table}

As discussed in the Introduction and shown in Fig.~\ref{fig:luminosity}, the neutrino emission in failed SNe starts at the time of the bounce of the outer core off the inner one and ends sharply at the time of the BH formation. This is true also for the corresponding GW signal, see e.g. Fig.~3 of Ref.~\cite{Cerda-Duran:2013swa}. While the observation of the neutrino signal is guaranteed with current neutrino detectors up to the Magellanic Clouds, the observation of the GW counterpart is still challenging \cite{Szczepanczyk:2021bka}. 
As demonstrated in previous works~\cite{Nakamura:2016kkl,Halim:2021yqa}, the use of a fixed temporal window to look for a GW burst and the combination of several detectors in temporal coincidence to decrease the false alarm rate probability, are powerful multi-messenger strategies to increase the GW detection efficiency. 

In this work we quantify the capability of the veto regions of LNGS detectors to provide the universal time of the BH formation for GW search ($T^{\text{\tiny GW}}_{\text{\tiny BH}}$). 
We perform 100 Monte Carlo (MC) simulations for a failed SN at $10$~kpc with both emission models discussed above.
For each detector we estimate the relative time of the first event $t^1=T^1-T^{\nu}_{\text{\tiny bounce}}$, where $T^1$ is the time of the first non-isolated event, the relative time of the last event $t^{\text{\tiny last}}=T^{\text{\tiny last}}-T^{\nu}_{\text{\tiny bounce}}$, where $T^{\text{\tiny last}}$ is the time of the last event, and the average rate of observed events $\xi=N_{\text{\tiny IBD}}/(t^{\text{\tiny last}}-t^{1})$. In Tab.~\ref{tab:results} we show the average values for $t^1$ and $t^{\text{\tiny last}}$ and their statistical fluctuations obtained from the MC simulations.

We are interested to estimate the time of BH formation at a GW detector related to the time of BH formation at the neutrino detector $T^{\nu}_{\text{\tiny BH}}$ through the equation $T^{\text{\tiny GW}}_{\text{\tiny BH}}=T^{\nu}_{\text{\tiny BH}}\pm t_{\text{\tiny fly}}$, where $t_{\text{\tiny fly}}$ is the time of fly between the neutrino and the GW detectors. In addition, we are also interested to provide the uncertainty $\delta T^{\text{\tiny GW}}_{\text{\tiny BH}}$ on this quantity.

The first proxy of the quantity $T^{\nu}_{\text{\tiny BH}}$ will be the time of the last neutrino event $T^{\text{\tiny last}}$. However, we find, in agreement with the results from Ref.~\cite{Beacom:2000qy}, that this time is systematically smaller than the BH formation time by an interval time that is well described by the data-driven quantity $1/\xi$, reported in the last column of Tab.~\ref{tab:results}. 
In addition, the same quantity can be assumed as a conservative estimation of the associated uncertainty $\delta T^{\nu}_{\text{\tiny BH}}$. In other words, for each neutrino detector the best proxy for the BH formation time is  
\begin{eqnarray}
T^{\nu}_{\text{\tiny BH}}&=&T^{\text{\tiny last}}+1/\xi\,,\\
\delta T^{\nu}_{\text{\tiny BH}}&=&1/\xi,
\end{eqnarray}  
where each term can be obtained by the observed events without any theoretical assumption on the emission model. 
Note that the data-driven quantity $\xi$ is also a powerful parameter to discriminate real SN signals from background and to increase the detection capability of neutrino detectors for small statistics signals~\cite{Casentini:2018bdf}. 

\begin{figure}
\centering
\includegraphics[width=0.49\textwidth]{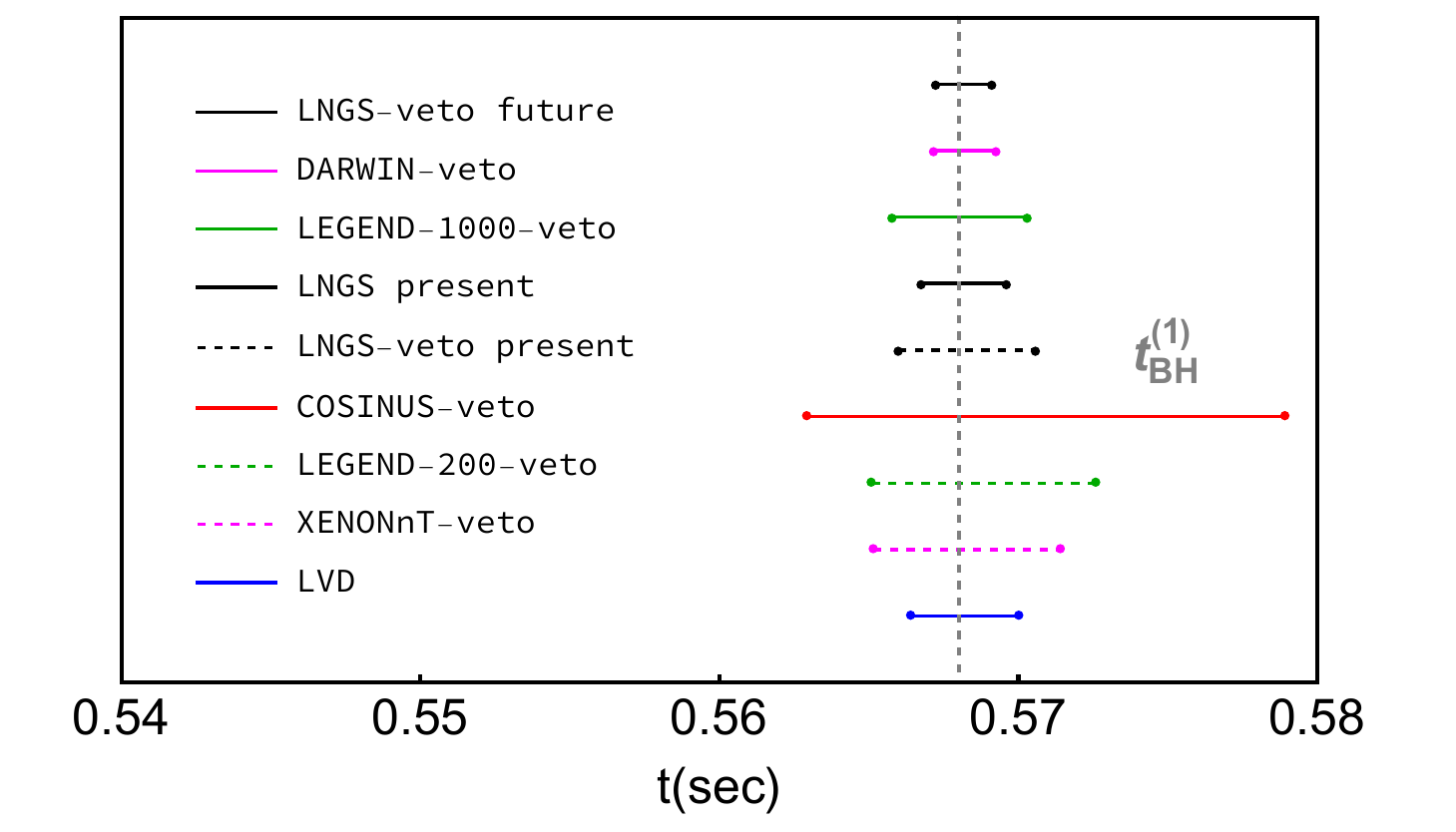}
\includegraphics[width=0.48\textwidth]{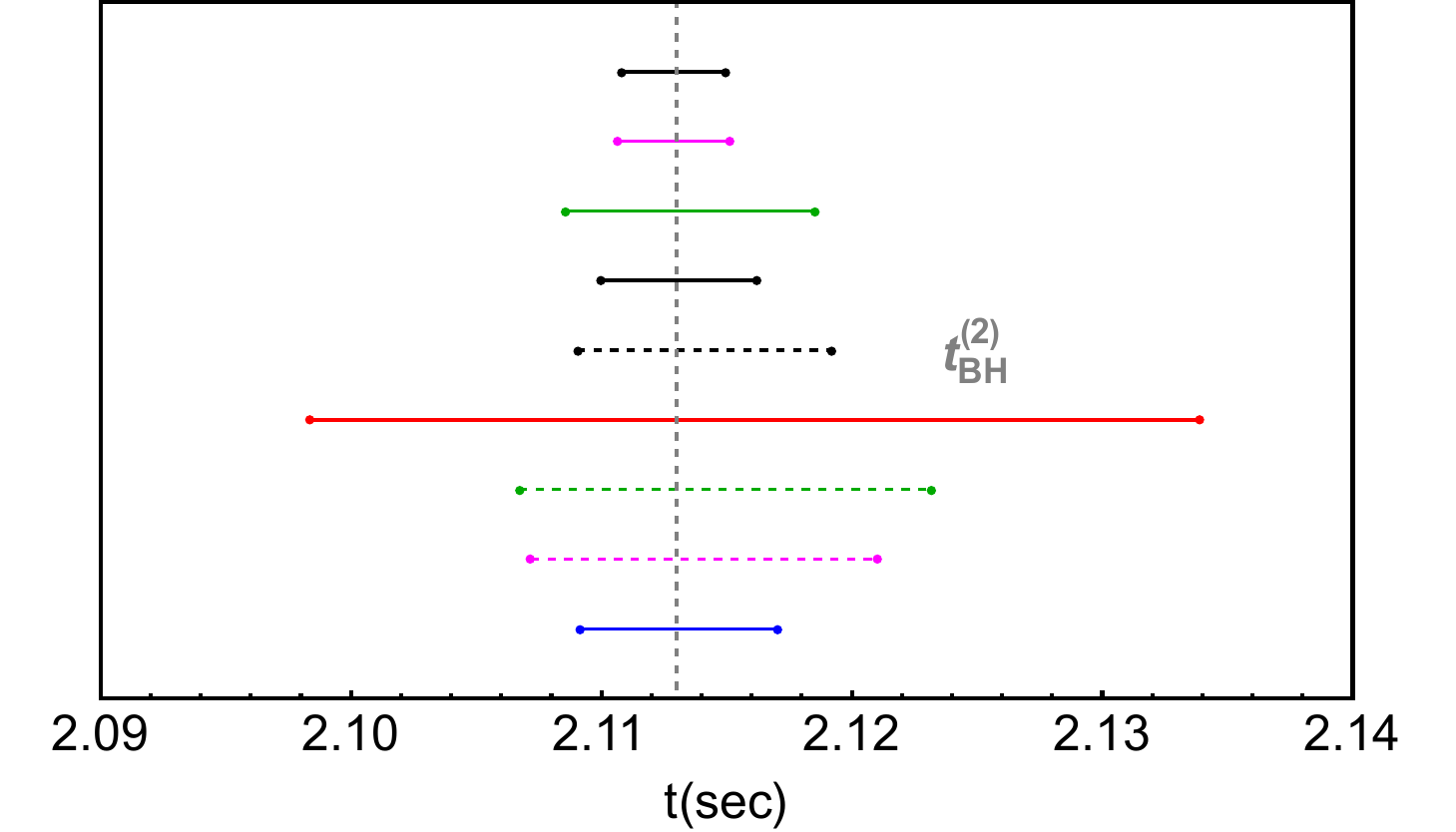}
\caption{The capability to identify the BH formation time of the different detectors considered in this paper for emission Model 1 (left panel) and emission Model 2 (right panel).}
\label{fig:BH_m1}
\end{figure}

In Fig.~\ref{fig:BH_m1} we show the obtained values of $T^{\nu}_{\text{\tiny BH}}\pm \delta T^{\nu}_{\text{\tiny BH}}$ for each neutrino detector considered in this work. Evidently, the real BH formation time is always within the reconstructed interval provided by these estimations.
The veto regions of the LNGS detectors can also be combined as modules of a bigger detector. In this case, the best proxy for the BH formation time will be 
\begin{eqnarray}
T^{\nu}_{\text{\tiny BH}}&=&\text{Max}[T^{\text{\tiny last}}_i]+1/\xi_{\text{Max}}\\
\delta T^{\nu}_{\text{\tiny BH}}&=&\sqrt{1/\sum_i\left(\xi_i^2\right)}
\end{eqnarray} 
where $i$ runs over all the considered detectors and $\xi_{\text{Max}}$ is the $\xi$ associated to $\text{Max}[T^{\text{\tiny last}}_i]$. In this case the uncertainty profits from the network configuration. 
The uncertainties on the black hole formation time obtained for the individual detectors are summarized in the last column of Tab.~\ref{tab:results} and range from 1~ms to 8~ms for emission Model 1, and from 2~ms to 18~ms for emission Model 2. Using the combined network of current veto regions the uncertainty can be improved to $\delta T^{\text{\tiny LNGS-veto}}_{\text{\tiny BH}}=2.3$~ms ($4.8$~ms) for Model 1 (Model 2).
These results can be compared with the sensitivity of LVD, namely $\delta T^{\text{\tiny LVD}}_{\text{\tiny BH}}=1.8$~ms ($3.9$~ms) showing that the veto region network could provide a similar precision as a dedicated Supernova experiment. Of course, at the present, all these detectors are in operation at LNGS and the overall sensitivity, "LNGS-present" in Fig.\ref{fig:BH_m1}, can reach even $\delta T^{\text{\tiny LNGS}}_{\text{\tiny BH}}=1.4$~ms ($3$~ms).
In the future, bigger detectors are expected to be hosted at LNGS. LEGEND-1000 can determine the time of BH formation within $\delta T^{\text{\tiny Legend-1000}}_{\text{\tiny BH}}=2$~ms ($5$~ms) while DARWIN could shrink the uncertainty to $\delta T^{\text{\tiny Darwin}}_{\text{\tiny BH}}=1.0$~ms ($2.2$~ms). When the possibility to combine these two detectors is considered we obtain the best sensitivity of $\delta T^{\text{\tiny LNGS}}_{\text{\tiny BH}}=0.9$~ms ($2.0$~ms).

Note that these detectors are not competitive with other observatories, e.g. Super-Kamiokande, in terms of providing the most precise measurement of $T^{\nu}_{\text{\tiny BH}}$, which can be up to one order of magnitude more precise than the LNGS measurement, see the discussion below. However, LNGS will provide important temporal information to the nearby GW detector Virgo and the future Einstein Telescope.   
Indeed, as stated before, the time of BH formation for the GW detector is given by $T^{\text{\tiny GW}}_{\text{\tiny BH}}=T^{\nu}_{\text{\tiny BH}} \pm t_{\text{\tiny fly}}$, where $t_{\text{\tiny fly}}$ is the time of fly between the detectors, $t_{\text{\tiny fly}}=\vec{r} \cdot \hat{n}$, where $\vec{r}$ is the distance-vector between the detectors and $\hat{n}$ is the direction pointing to the Supernova. %
The associated uncertainty is $\delta T^{\text{\tiny GW}}_{\text{\tiny BH}}=\delta T^{\nu}_{\text{\tiny BH}} +\delta t_{\text{\tiny fly}}$. The error $\delta t_{\text{\tiny fly}}$ is negligible for an astronomically identified SN. However, it can be equal to $t_{\text{\tiny fly}}$ for an unknown SN position. For a failed SN the electromagnetic counterpart is missing and the identification of the SN position in the sky is more challenging.
The distance between LNGS and Virgo or the Einstein telescope is very small so that $t_{\text{\tiny fly}}\leq 1$ ms. This means that, even in the worse scenario of slow BH formation (emission Model 2), the temporal uncertainty that the current network of LNGS detectors can provide to Virgo is 
\begin{equation}
    \delta T^{\text{\tiny GW}}_{\text{\tiny BH}}=\delta T^{\text{\tiny LNGS}}_{\text{\tiny BH}}+\delta t_{\text{\tiny fly}}=4~\text{ms}\,.  
\end{equation}
We can compare this quantity with the equivalent one obtained for Super-Kamiokande with $32$~kton of water and an energy threshold of $E_{\text{\tiny thr}}=7$~MeV. Super-Kamiokande will collect a very big number of events (13776) with a corresponding very small uncertainty $\delta t^{\text{\tiny SK}}_{\text{\tiny BH}}=0.15$~ms, in agreement with a previous estimation in Ref.~\cite{Brdar:2018zds}.
However, the distance between Super-Kamiokande and Virgo or the Einstein Telescope is quite large so that $t_{\text{\tiny fly}}$ could be as large as $28$~ms providing a final uncertainty window of 

\begin{equation}
    \delta T^{\text{\tiny GW}}_{\text{\tiny BH}}=\delta T^{\text{\tiny SK}}_{\text{\tiny BH}}+\delta t_{\text{\tiny fly}}=28.3~\text{ms}
\end{equation}
for the same emission Model 2. Of course the uncertainty on the time of fly could be reduced by exploiting the information provided by directional neutrino interactions as the elastic scattering on electrons~\cite{Tomas:2003xn} or by triangulation between different neutrino detectors~\cite{Coleiro:2020vyj} as discussed also for this specific case in Ref.~\cite{Sarfati:2021vym}. However, all these techniques require deep analysis of neutrino data in order to tag the IBD channel~\cite{Marti:2019dof} or to combine the neutrino data streams of different detectors around the world~\cite{Muhlbeier:2013gwa}, while our information can be provided immediately to the GW detector as soon as the neutrino data are collected.

\section{Conclusions}
\label{sec:conclusions}
In conclusion, we have investigated the role of veto regions of detectors mainly dedicated to other physics topics, to detect supernova neutrinos. We concentrate on the peculiar case of the LNGS infrastructure where three different detectors (COSINUS, LEGEND-200 and XENONnT) use water tanks of different shapes and volumes as veto regions, that can be easily combined to work together at a network of small detectors. 
We show that this combination of small free-of-charge Cherenkov detectors is sensitive to neutrinos from CCSNe up to the Magellanic Clouds. In particular, we quantify the capability of these detectors to identify the time of BH formation for a failed SN. 
This information can be very relevant to look for the gravitational wave counterpart to the neutrino emission. We show that the sensitivity of current LNGS detectors working as a network could  provide the time of BH formation with the smallest uncertainty thanks to the very small distance between the LNGS facility and the Virgo gravitational wave interferometer.
We also provide a projection of this sensitivity with future detectors expected in the same laboratory. We want to stress that the approach described in this work is completely model independent and can be applied to any type of BH forming SN, also the ones that collapse to a BH after the explosion~\cite{Burrows:2023nlq}. 
This statement is not true, however, if we were interested to identify the starting time of the neutrino signal, i.e. the time of the bounce. In this case the estimation of the $T^{\nu}_{\text{\tiny bounce}}$ will be affected by a model dependent uncertainty, that is quantified as the absolute value of $t^{1}$ in our Tab.~\ref{tab:results}. 
This term, usually called response time of the detector~\cite{Pagliaroli:2009qy} cannot be deduced by the neutrino data-stream without any assumption on the emission model.   

Finally, as a last remark we point out that our analysis is very conservative and timely. It is
conservative because it provides a lower limit of the total sensitivity of the LNGS detectors. 
Indeed we are considering only the IBD detection channel and only the veto regions of the detectors. The sensitivity could be improved by including other detection channels both in water and in liquid scintillator and by considering also the inner regions of the dark matter or $0\nu\beta\beta$ detectors that are also sensitive to SN neutrinos~\cite{Lang:2016zhv}. The corresponding increase of statistics can be easily converted in a reduced temporal uncertainty on the time of BH formation. The analysis is timely because the COSINUS and LEGEND-200 detectors are now still defining the best configuration for their veto regions.

\acknowledgements 
The authors thank N. Di Marco for useful discussions. The work of G.P. is partially supported by  grant number 2022E2J4RK "PANTHEON: Perspectives in Astroparticle and
Neutrino THEory with Old and New messengers" under the program PRIN 2022 funded by the Italian Ministero dell’Universit\`a e della Ricerca (MUR) and by the European Union – Next Generation EU.


%

\end{document}